# Intrinsic 2D Ferromagnetism in $V_5Se_8$ Epitaxial Thin Films

*Masaki Nakano,*[\*,†,‡,¶] *Yue Wang,*[†,¶] *Satoshi Yoshida,*[†,¶] *Hideki Matsuoka,*[†] *Yuki Majima,*[†] *Keisuke Ikeda,*[§] *Yasuyuki Hirata,*[§,∥] *Yukiharu Takeda,*[⊥] *Hiroki Wadati,*[§,∥,★] *Yoshimitsu Kohama,*[∥] *Yuta Ohigashi,*[†] *Masato Sakano,*[†] *Kyoko Ishizaka,*[†,‡] *and Yoshihiro Iwasa*[†,‡]

[†]Quantum-Phase Electronics Center and Department of Applied Physics, the University of Tokyo, Tokyo 113-8656, Japan.

[‡]RIKEN Center for Emergent Matter Science (CEMS), Wako 351-0198, Japan.

[§]Department of Physics, the University of Tokyo, Tokyo 113-0033, Japan.

[∥]The Institute for Solid State Physics, the University of Tokyo, Kashiwa 227-8581, Japan.

[⊥]Materials Sciences Research Center, Japan Atomic Energy Agency, Hyogo 679-5148, Japan.

**ABSTRACT**: The discoveries of intrinsic ferromagnetism in atomically-thin van der Waals crystals have opened up a new research field enabling fundamental studies on magnetism at two-dimensional (2D) limit as well as development of magnetic van der Waals heterostructures. To date, a variety of 2D ferromagnetism has been explored mainly by mechanically exfoliating 'originally ferromagnetic (FM)' van der Waals crystals, while bottom-up approach by thin film growth technique has demonstrated emergent 2D ferromagnetism in a variety of 'originally non-FM' van der Waals materials. Here we demonstrate that $V_5Se_8$ epitaxial thin films grown by molecular-beam epitaxy (MBE) exhibit emergent 2D ferromagnetism with intrinsic spin polarization of the V $3d$ electrons despite that the bulk counterpart is 'originally antiferromagnetic (AFM)'. Moreover, thickness-dependence measurements reveal that this newly-developed 2D ferromagnet could be classified as an itinerant 2D Heisenberg ferromagnet with weak magnetic anisotropy, broadening a lineup of 2D magnets to those potentially beneficial for future spintronics applications.

Exploring novel 2D magnets that exhibit long-range magnetic order at the 2D limit is one of central topics in modern condensed-matter researches based on 2D materials and their heterostructures.[1-5] There, the most-frequently-used approach is to mechanically exfoliate FM van der Waals crystals down to the ultrathin regime, providing a simple but versatile route to exploration of intrinsic 2D ferromagnetism as initially demonstrated on insulating ferromagnets, $Cr_2Ge_2Te_6$ (CGT) and $CrI_3$.[1,2] Those pioneering works as well as the recent study on metallic ferromagnet, $Fe_3GeTe_2$ (FGT),[3] have experimentally verified the fundamental aspect of 2D ferromagnets at two extreme conditions; long-range FM order could be stabilized down to the monolayer limit in Ising-type ferromagnets with strong uniaxial magnetic anisotropy (*i.e.*, $CrI_3$ and FGT), whereas it is strongly suppressed at the 2D limit in Heisenberg-type ferromagnets with weak anisotropy (*i.e.*, CGT) due to thermal fluctuations in accord with the Mermin-Wagner theorem.[6] However, very intriguingly, it turned out that 2D Heisenberg ferromagnets give rise to an emergent functionality enabling control of the FM transition temperature by external magnetic fields,[1] which is unique to this class of 2D ferromagnets with weak magnetic anisotropy.

Apart from mechanical exfoliation, recent development of thin film growth technique has enabled us to investigate magnetic properties of a wide variety of van der Waals materials down to the monolayer limit even including hardly-cleavable and thermodynamically-metastable compounds, providing a unique and powerful path to exploration of novel 2D magnets. In fact, those atomically-thin samples grown by bottom-up approach have often provided emergent room-temperature 2D ferromagnetism in originally non-FM compounds as demonstrated in monolayer $VSe_2$ and $MnSe_x$ epitaxial thin films grown by MBE as well as in $VTe_2$ and $NbTe_2$ ultrathin nanoplates grown by chemical-vapor deposition.[7-9] However, in those studies, magnetic properties were mainly characterized by magnetization measurements using a superconducting quantum interference device magnetometer, which could not distinguish intrinsic FM signals from extrinsic ones in particular when the total magnetic moments were very small in the ultrathin regime. Recent spectroscopic studies on MBE-grown monolayer VSe$_2$ suggest that the ground state of this compound should be the charge-density wave state rather than the FM state,[10-14] indicating the essential importance of evaluation of magnetic properties by different approaches.

In this study, we examined magnetic properties of vanadium selenide epitaxial thin films grown by our MBE process[15-17] through anomalous Hall effect (AHE) and x-ray magnetic circular dichroism (XMCD) measurements, which should enable us to probe only intrinsic FM signals. Importantly, detailed structural characterization verified that our films evolve from VSe$_2$ to V$_5$Se$_8$ soon after the formation of the first layer. Figure 1a,b illustrates a schematic structure of V$_5$Se$_8$ crystal,[18] which is characterized by the 2D VSe$_2$ sheets separated by the layers of the periodically-aligned V atoms (labelled as V$^I$). Figure 1c shows the cross-sectional scanning transmission electron microscope image of our typical film, indicating the existence of the well-ordered V$^I$ atoms in between the VSe$_2$ layers. We confirmed that those V interlayers were formed just after the formation of the first VSe$_2$ layer. Figure 1d shows the reflection high energy electron diffraction (RHEED) pattern taken at the end of the growth, and Figure 1e exhibits the evolution of the in-plane lattice parameter during the growth evaluated from the spacing between the RHEED streaks. The lattice parameter initially matched the $a$-axis length of VSe$_2$,[19] while it largely deviated from that of VSe$_2$ during the formation of the second VSe$_2$ layer and stayed almost unchanged until the end of the growth. The saturated value tuned out to match the $a^*$-axis length of V$_5$Se$_8$,[18] indicating the formation of V$_5$Se$_8$ after the second layer. The formation of V$_5$Se$_8$ lattice could be also confirmed by the out-of-plane x-ray diffraction pattern shown in Figure 1f, indicating that the out-of-plane lattice parameter was close to the $c^*$-axis length of V$_5$Se$_8$ rather than the $c$-axis length of VSe$_2$.[18,19]

Bulk V$_5$Se$_8$ is known to be an itinerant antiferromagnet with the Néel temperature $T_N$ ~ 30 K,[20,21] while our V$_5$Se$_8$ epitaxial thin films turned out to exhibit the spontaneous FM order that is missing in bulk. Figure 2a shows the Hall-effect data of our 30 ML-thick film, demonstrating negative AHE signals with clear magnetic hysteresis loop at $T$ = 2 K. This hysteresis disappeared with increasing temperature at around $T$ ~ 10 K, while the AHE signals survived up to $T$ ~ 20 K. These results verified that our film entered the FM phase below $T$ ~ 20 K while the spontaneous FM order developed below $T$ ~ 10 K. On the other hand, the magnetoresistance (MR) measurements proved the existence of the AFM order together with the FM order. The previous studies on a sister sulfide compound V$_5$S$_8$ demonstrated unique anisotropy in MR in the AFM phase.[22,23] Figure 2b shows the MR curves of our 30 ML-thick film with the out-of-plane and in-plane magnetic fields taken at different temperatures. Negative MR was observed in both configurations, while the magnitude was different depending on the direction of the magnetic fields, which should be associated with the existence of the AFM order.[22,23] The onset temperature was determined to be around $T$ = 30 K, which is consistent to $T_N$ of bulk V$_5$Se$_8$ and slightly above the onset temperature of the FM order. These results therefore suggest that our film hosts the AFM phase with $T_N$ ~ 30 K together with the FM phase below $T$ ~ 20 K.

Thickness dependence measurements revealed that the observed spontaneous FM order was strongly suppressed at the monolayer limit. Figure 3a shows the Hall-effect data at $T$ = 2 K for the different-thickness samples, the 30 ML-, 8 ML-, and 3 ML-thick films. All the films demonstrated clear AHE signals, while the hysteresis loop associated with the spontaneous FM order nearly disappeared for the 3 ML-thick film. Considering that the 3 ML-thick film should have the same structure and composition as those of the thicker films (see Figure 1e and Supporting Information), the observed suppression of the spontaneous FM order at the 2D limit should be attributed to the dimensionality effect. Figure 3b shows the results of the detailed temperature-dependence measurements, where the anomalous Hall resistivity ($\rho_{AH}$) at the magnetic field $\mu_0 H = -3$ T ($\rho_{AH, sat}$, corresponding to the saturation magnetization) and at $\mu_0 H = 0$ T ($\rho_{AH, rem}$, corresponding to the remanent magnetization) are both plotted as a function of temperature (see Supporting Information for the corresponding AHE data). We here define $T_C^*$ and $T_C$ as the temperatures at which $\rho_{AH, sat}$ and $\rho_{AH, rem}$ start to develop, respectively, which were determined to be 17 K and 7 K for the 30 ML-thick film. With reducing thickness, on the other hand, $T_C$ turned out to be significantly suppressed down below 2 K for the 3 ML-thick film, while $T_C^*$ remained almost unchanged. Those results are essentially the same as those observed in CGT nano-thick crystals,[1] where the spontaneous FM order was suppressed at the monolayer limit, which is a characteristic feature of Heisenberg-type ferromagnets with weak magnetic anisotropy. Taken together, we conclude that this newly-developed ferromagnet should be classified as an itinerant 2D Heisenberg ferromagnet, a new type of 2D magnets that have not been developed so far including exfoliated flakes. Figure 3c summarizes the evolutions of $T_C^*$ and $T_C$ as a function of the layer number together with that of the $\rho_{AH}/\rho_{xx}$ value calculated with $\rho_{AH, sat}$ at $T = 2$ K.

The observed AHE signals verified the existence of the spin-polarized carriers in our $V_5Se_8$ epitaxial thin films, while XMCD measurements elucidated that the emergent FM phase in fact arises from intrinsic spin polarization of the V $3d$ electrons. Owing to the element-specific nature as well as the very high sensitivity, XMCD has been widely used to investigate the origin of the magnetism even in diluted systems as well as interface systems.[24,25] Figure 4a shows the x-ray absorption spectra (XAS) of the 30 ML-thick film near the V $L_{2,3}$-edges taken at the base temperature $T$ = 5 K with right-handed circularly-polarized (RCP) and left-handed circularly-polarized (LCP) incident light under the out-of-plane magnetic fields. The helicity switching of circularly-polarized light was operated at 1 Hz using the twin helical undulator[26]. There was clear difference in the XAS spectra between RCP and LCP setups, which changed its sign under reversal of the magnetic field direction, indicating the existence of the intrinsic XMCD signals. The resultant XMCD spectra are shown in Figure 4b, demonstrating that the finite XMCD signals appeared only near the absorption peaks. This indicates that those signals are indeed originating from intrinsic spin polarization of the V 3$d$ electrons. Figure 4c shows the magnetic field dependence of the XMCD intensity taken at the photon energy of $h\nu$ = 517.8 eV, clearly demonstrating non-linear behavior, which should be arising from the FM order. This non-linear behavior turned out to disappear at $T$ = 25 K, suggesting that the onset temperature of the FM order should be located between $T$ = 5 K and $T$ = 25 K, which is consistent to the transport results with $T_C^*$ = 17 K for the 30 ML-thick film. The linear response observed at $T$ = 25 K should arise either from the AFM phase or from the enhanced paramagnetic response near $T_C^*$, although further studies are needed to address its origin. Nevertheless, the obtained results strongly support that the emergent FM phase in our V$_5$Se$_8$ epitaxial thin films should have intrinsic origin arising from spin polarization of the V 3$d$ electrons.

As a driving force that could stabilize the FM phase in V$_5$Se$_8$ epitaxial thin films, we here propose three possible origins. Considering that bulk V$_5$Se$_8$ hosts the AFM order and that the FM order in our V$_5$Se$_8$ films appears just below $T_N$, it is natural to think that the observed ferromagnetism could be categorized as the weak ferromagnetism (or the canted antiferromagnetism in other words), which should originate from the Dzyaloshinskii-Moriya (DM) interaction induced by the broken space inversion symmetry. Given that V$_5$Se$_8$ is a centrosymmetric material,[18] the DM interaction should appear as a consequence of the local symmetry breaking at the substrate-film interface. A second possibility could be the effect of the epitaxial strain. The in-plane lattice parameters of our $V_5Se_8$ films evaluated by the RHEED analysis were somewhat larger than that of bulk $V_5Se_8$ (see Figure 1e), implying the existence of the tensile strain in our $V_5Se_8$ films. This suggests that the distance between the interlayer $V^I$ atoms is a bit increased, which might influence the magnetic exchange interaction in $V_5Se_8$. Another possibility could be the dimensionality effect. The recent study on a sister sulfide compound $V_5S_8$ reported a signature of the evolution of the magnetic ground state from the AFM phase to the FM phase with reducing thickness, which was attributed to the dimensionality effect.[27] Although the material is different, a similar effect might be relevant to our $V_5Se_8$ films. And also, this implies that reducing dimensionality not only leads to the suppression of the spontaneous FM order at the 2D limit, but also triggers the evolution of the magnetic ground state at the relatively thicker regime in this material family. Those three mechanisms are just examples, and further studies are required to address the underlying microscopic mechanism both from theory and experiment viewpoints. However, we note that the AFM phase and the FM phase should be almost energetically degenerate in this material family due to its unique magnetic structure, and therefore, a small modification in their structural and electronic properties could trigger the evolution of the magnetic ground state. Detailed discussions on the magnetic structure are provided in Supporting Information.

The present study demonstrates the emergence of intrinsic ferromagnetism in $V_5Se_8$ epitaxial thin films grown by MBE. The thickness dependent AHE verified that this material system forms a new class of 2D magnets categorized as an itinerant 2D Heisenberg ferromagnet, which has not been developed so far including exfoliated flakes. These Heisenberg-type ferromagnets are characterized with weak magnetic anisotropy, which should be advantageous for controlling magnetic properties by external means including magnetic and electric fields as well as by the proximity effect in heterostructures, potentially providing a novel functionality unachievable with Ising-type ferromagnets with strong uniaxial magnetic anisotropy. The current work also demonstrates the fundamental importance of employing AHE and XMCD for evaluation of intrinsic ferromagnetism in 2D materials in the ultrathin regime.

**ASSOCIATED CONTENT**

**Supporting Information**

The Supporting Information is available free of charge on the ACS Publications website at DOI:

Experimental details, magnetic structure of $V_5S_8$, further characterization of the AFM phase, and the detailed transport data including the temperature dependence of the resistivities and the Hall coefficients as well as the detailed AHE data of the different-thickness samples. (PDF)

**AUTHOR INFORMATION**

**Corresponding Author**

*E-mail: nakano@ap.t.u-tokyo.ac.jp

**Present Addresses**

★Graduate School of Material Science, University of Hyogo, Hyogo 678-1297, Japan.

**Author Contributions**

¶M.N., Y.W., and S.Y. contributed equally to this work. Y.W., S.Y., H.M., and Y.M. grew and characterized the thin films. M.N., Y.W., H.M., and Y.M. performed transport measurements and analyzed the data. M.N., S.Y., K.Ikeda, Y.H., Y.T., and H.W. performed x-ray magneto-optical measurements. M.N., H.M., and Y.K. performed magneto-transport measurements at high fields. Y.O. and M.S. supported the experiments. M.N., K.Ishizaka, and Y.I. supervised the study. M.N. wrote the manuscript. All the authors discussed the results and commented on the manuscript.

**Notes**

The authors declare no competing financial interest.

**ACKNOWLEDGMENT**

We are grateful to S. M. Bahramy for valuable discussions and to Y. Kashiwabara, T. Shitaokoshi, K. Matsui, M. Kawasaki, M. Uchida, and T. Fujita for experimental help. This work was supported by Grants-in-Aid for Scientific Research (Grant Nos. 19H05602, 19H02593, and 19H00653) and A3 Foresight Program from the Japan Society for the Promotion of Science (JSPS). M.N. was partly supported by The Murata Science Foundation. H.M. was supported by JSPS through Program for Leading Graduate Schools (MERIT). The synchrotron x-ray magneto-optical measurements were performed at BL23SU in SPring-8 (Proposal No. 2018A3843).

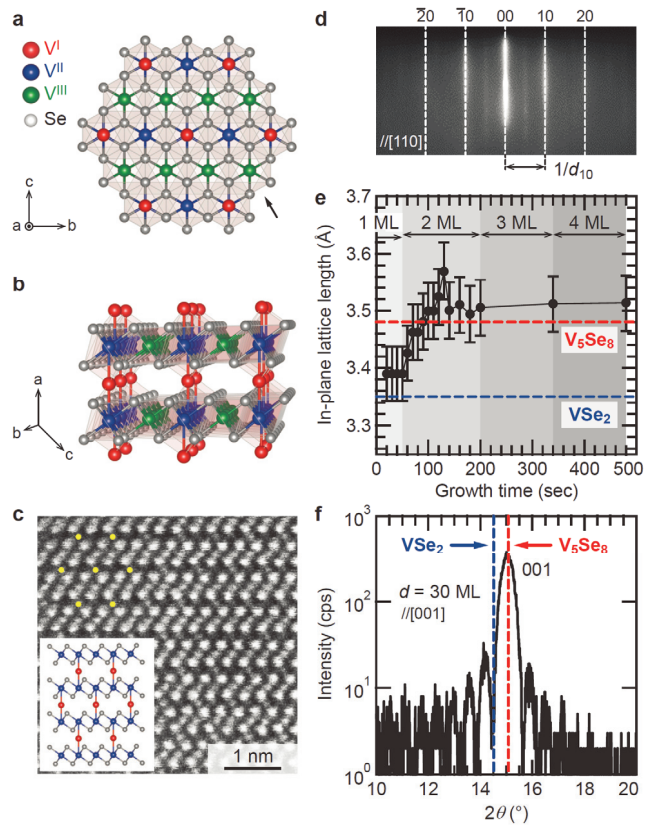

**Figure 1.** Basic structural properties of $V_5Se_8$ epitaxial thin films grown by MBE. (a) Schematic top view and (b) side view of $V_5Se_8$ crystal. (c) The cross-sectional scanning transmission electron microscope image of the 30 ML-thick film taken along $[\bar{1}10]$ azimuth of the substrate (the direction indicated by an arrow in a). The yellow dots indicate the positions of the interlayer V atoms. The inset shows the corresponding schematic crystal structure. (d) The RHEED pattern of a typical film taken along [110] azimuth of the substrate after the whole growth process. The Miller indices of the film are converted from original ones characterized with $a$-, $b$-, and $c$-parameters defined in a and b to the setting of $VSe_2$ lattice with $a^*$-, $b^*$-, and $c^*$-parameters, which were calculated using the relations, $a^* = b/2$, $b^* = (c/2)\sin\beta \times 2/\sqrt{3}$, and $c^* = a/2$. (e) The evolution of the in-plane lattice parameter during the film growth calculated from the spacing between the RHEED streaks. The layer number is defined as the number of the host $VSe_2$ layer. The $a$-axis length of $VSe_2$ and the $a^*$-axis length of $V_5Se_8$ are taken from the literatures.[18,19] (f) The out-of-plane x-ray diffraction pattern of the 30 ML-thick film. The expected peak positions of $VSe_2$ and $V_5Se_8$ are calculated from the $c$-axis length of $VSe_2$ and the $c^*$-axis length of $V_5Se_8$ taken from the literatures.[18,19]

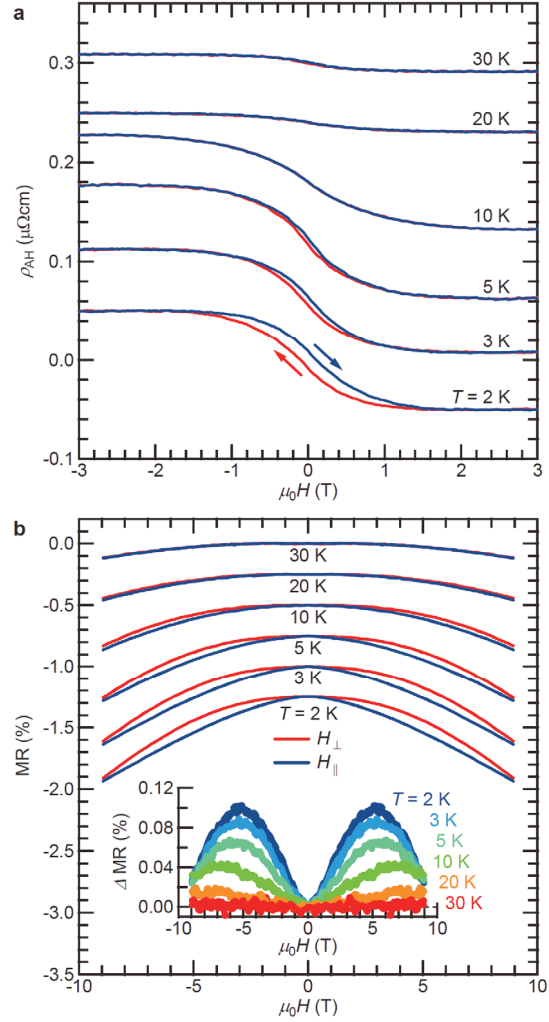

**Figure 2.** The magneto-transport properties of the 30 ML-thick V$_5$Se$_8$ epitaxial thin film. (a) $\rho_{AH}$ of the 30 ML-thick film calculated with the anti-symmetrized transverse resistivity as a function of the magnetic fields at different temperatures. The data at $T$ = 30 K, 20 K, 10 K, 5 K, and 3 K are vertically shifted by 0.30, 0.24, 0.18, 0.12, and 0.06 μΩcm, respectively. (b) The symmetrized MR curves of the 30 ML-thick film with the out-of-plane (red) and in-plane (blue) magnetic fields at different temperatures. The MR value is defined as $[\rho(\mu_0 H)-\rho_0]/\rho_0$. The data at $T$ = 20 K, 10 K, 5 K, 3 K, and 2 K are vertically shifted by -0.25, -0.50, -0.75, -1.00, and -1.25 %, respectively. The inset shows the difference in MR defined as the MR with the out-of-plane fields minus that with the in-plane fields at different temperatures.

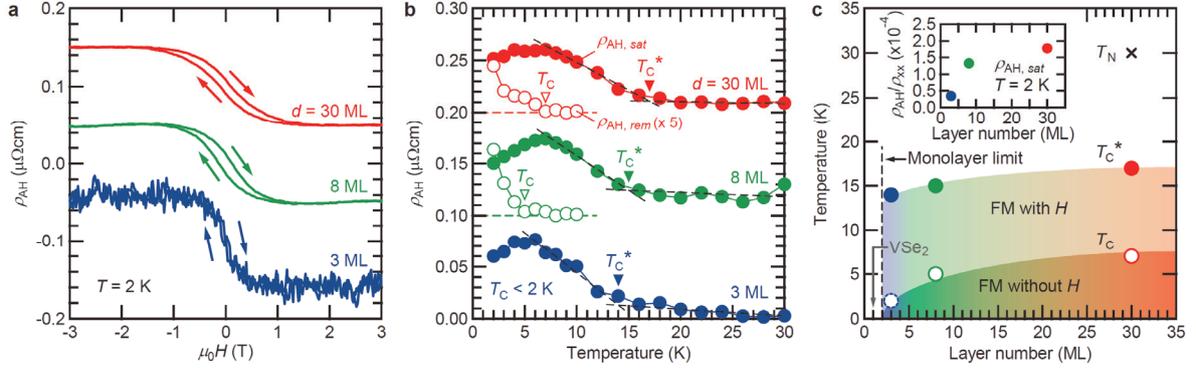

**Figure 3.** Dimensionality effect on ferromagnetism of $V_5Se_8$ epitaxial thin films down to the monolayer limit. (a) The $\rho_{AH}$ versus $\mu_0 H$ curves of the 30 ML- (red), 8 ML- (green), and 3 ML- (blue) thick films at $T = 2$ K. The data of the 30 ML- and 3 ML-thick films are vertically shifted by 0.1 and -0.1 µΩcm, respectively. The data of the 30 ML-thick film is the same as that shown in Figure 2a. (b) The temperature dependences of $\rho_{AH}$ of the same samples. $\rho_{AH,\,sat}$ (filled symbols) corresponds to $\rho_{AH}$ at $\mu_0 H = -3$ T while $\rho_{AH,\,rem}$ (open symbols) corresponds to $\rho_{AH}$ at $\mu_0 H = 0$ T. The data of the 30 ML- and 8 ML-thick films are vertically shifted by 0.20 and 0.10 µΩcm, respectively. $T_C^*$ and $T_C$ are defined as the temperatures at which $\rho_{AH,\,sat}$ and $\rho_{AH,\,rem}$ start to develop, respectively. (c) The resultant phase diagram, where $T_C^*$ (filled symbols) and $T_C$ (open symbols) are plotted as a function of the layer number. The monolayer limit corresponds to the 2 ML-thick film, which consists of two $VSe_2$ sheets separated by one layer of the periodically-aligned V atoms. $T_N$ (black cross) of the 30 ML-thick film is also plotted. The inset shows the corresponding $\rho_{AH}/\rho_{xx}$ values of those films evaluated with $\rho_{AH,\,sat}$ at $T = 2$ K.

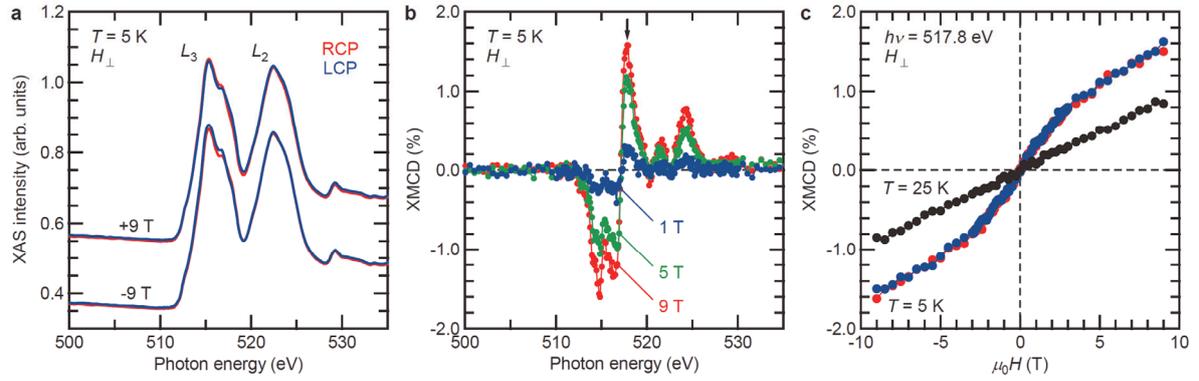

**Figure 4.** The x-ray magneto-optical properties of the 30 ML-thick $V_5Se_8$ epitaxial thin film. (a) The XAS of the 30 ML-thick film near the V $L_{2,3}$-edges taken under illuminations of RCP and LCP with the out-of-plane magnetic fields $\mu_0 H = \pm 9$ T at the base temperature $T = 5$ K. (b) The averaged XMCD spectra at the different magnetic fields. Each data was calculated by subtracting the $I_{RCP}$-$I_{LCP}$ data taken at the positive magnetic fields from those at the negative fields. (c) The magnetic field dependences of the anti-symmetrized XMCD intensity taken at the photon energy of $h\nu = 517.8$ eV indicated by an arrow in b measured at $T = 5$ K (red and blue symbols for decreasing- and increasing-field sweeps, respectively) and at $T = 25$ K (black symbols for decreasing-field sweep).

**TOC FIGURE:**

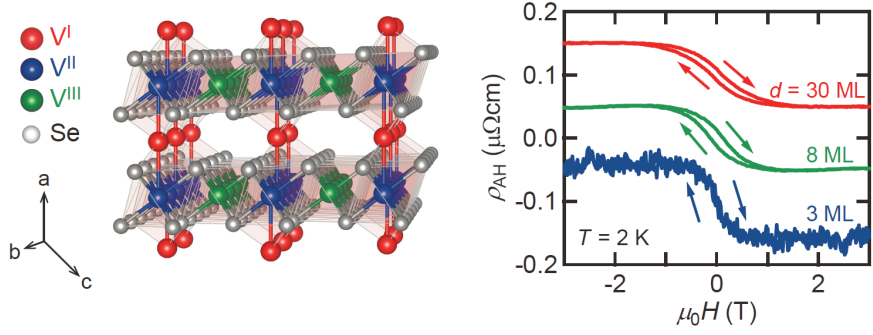

Supporting Information for

# Intrinsic 2D Ferromagnetism in $V_5Se_8$ Epitaxial Thin Films

*Masaki Nakano,*[*,†,‡,¶] *Yue Wang,*[†,¶] *Satoshi Yoshida,*[†,¶] *Hideki Matsuoka,*[†] *Yuki Majima,*[†] *Keisuke Ikeda,*[§] *Yasuyuki Hirata,*[§,∥] *Yukiharu Takeda,*[⊥] *Hiroki Wadati,*[§,∥,★] *Yoshimitsu Kohama,*[∥] *Yuta Ohigashi,*[†] *Masato Sakano,*[†] *Kyoko Ishizaka,*[†,‡] *and Yoshihiro Iwasa*[†,‡]

[†]Quantum-Phase Electronics Center and Department of Applied Physics, the University of Tokyo, Tokyo 113-8656, Japan.

[‡]RIKEN Center for Emergent Matter Science (CEMS), Wako 351-0198, Japan.

[§]Department of Physics, the University of Tokyo, Tokyo 113-0033, Japan.

[∥]The Institute for Solid State Physics, the University of Tokyo, Kashiwa 227-8581, Japan.

[⊥]Materials Sciences Research Center, Japan Atomic Energy Agency, Hyogo 679-5148, Japan.

## A. Experimental details

### Thin film growth and structural characterizations

All the films were grown on insulating $Al_2O_3$ (sapphire) (001) substrates by molecular-beam epitaxy (EIKO Engineering) by following our recently-established growth process.[15-17] The growth temperature was set to 450 ºC. During the growth, V was supplied by an electron beam evaporator with the evaporation rate of ~ 0.01 Å/s, while Se was supplied by a standard Knudsen cell throughout the growth process with the rate of ~ 2.0 Å/s. The obtained films were annealed *in-situ* at the same temperature for 30-60 minutes to improve the crystallinity. All the films were covered by thick insulating Se capping layers to protect their surfaces from oxidization. The crystallinities of the obtained films were characterized *in-situ* by reflection high energy electron diffraction and *ex-situ* by a four-circle x-ray diffractometer (PANalytical, X'Pert MRD). The local structure was characterized *ex-situ* by scanning transmission electron microscope (JEOL, JEM-ARM200F).

## Transport measurements

All the films were cut into Hall-bar shape by mechanical scratching over the Se capping layers in order to define the channel regions (typically a few hundreds of micrometers). The electrical transport properties were characterized by Physical Property Measurement System (Quantum Design, PPMS). The magnetoresistance curve measured up to $\mu_0 H \sim 60$ T was taken by the pulsed-field system at the Institute for Solid State Physics, the University of Tokyo.

## X-ray magnetic circular dichroism measurements

The synchrotron x-ray magneto-optical measurements including XAS and XMCD measurements were performed at BL23SU in SPring-8. The helicity switching of circularly-polarized light was operated at 1 Hz using the twin helical undulator.[26] All the data were taken in the total electron yield mode at different temperatures under different magnetic fields. The magnetic field direction was perpendicular to the sample and parallel to the incident beam direction.

## B. Magnetic structure of bulk $V_5X_8$ (X = S, Se)

A family of $V_5X_8$ (X = S, Se) has been studied for a long time since '70. Figure S1a shows a schematic structure of $V_5Se_8$.[18] There are three inequivalent V atoms drawn by three different colors. $V^I$ (red) represents the interlayer V atoms while $V^{II}$ (blue) and $V^{III}$ (green) correspond to the V atoms constituting the $VSe_2$ layers. The difference between $V^{II}$ and $V^{III}$ is whether the V atoms are bonded to the $V^I$ atoms or not. Magnetic properties of this material system have been well studied mainly on its sister sulfide compound $V_5S_8$ ($T_N \sim 25\text{-}35$ K),[21-23,28-32] showing that magnetism is hosted by the $V^I$ atoms while itinerancy is provided by the 2D $VS_2$ sheets containing the $V^{II}$ and $V^{III}$ atoms. Figure S1b illustrates the magnetic structure of $V_5S_8$ in the AFM phase determined by the neutron diffraction measurements.[32] The magnetic moments are thought to be arising mainly from the $V^I$ atoms, whereas those from the $V^{II}$ and $V^{III}$ atoms are believed to be negligible. The local spins at the $V^I$ atoms are ferromagnetically aligned along the out-of-plane direction, forming a 2D 'FM sheet' within the *a-b* plane. The neighboring two FM sheets form one 'FM block', which is antiferromagnetically coupled to the next FM block, forming a magnetic superstructure. This unique magnetic structure suggests that the FM coupling between two FM sheets and the AFM coupling between two FM blocks are almost energetically degenerate. The direction of each spin is a bit tilted from the *a*-axis direction toward the *c*-axis direction within the *a-c* plane by approximately 10 degree. Although the magnetic structure of $V_5Se_8$ has not yet been determined by neutron diffraction measurements, we assume it should be similar to that of $V_5S_8$ because of many similarities in terms of the crystal structure as well as the electronic and magnetic properties.

## C. Further characterizations of the AFM phase

Figure S2a shows the MR curve of another 30 ML-thick $V_5Se_8$ epitaxial thin film with the out-of-plane magnetic fields taken at $T = 2$ K. There was small but finite magnetic hysteresis loop observed near $\mu_0 H = \pm 5$ T, most likely originating from the spin-flop transition in the AFM phase.[22,23]. Figure S2b presents the MR curve measured up to $\mu_0H \sim 60$ T with the out-of-plane magnetic fields at $T = 0.7$ K by the pulsed-field system. The negative MR saturated at around $\mu_0H \sim 50$ T, which is consistent to the saturation of the magnetization measured with bulk $V_5Se_8$ polycrystalline samples,[21] indicating that the negative MR originates from the AFM order. The Zeeman energy at $\mu_0H = 50$ T corresponds to the thermal energy at $T \sim 30$ K, which roughly matches $T_N$ of our 30 ML-thick film characterized by the MR anisotropy measurements (see the main text for details). Figure S3a-f shows the MR curves of the 30 ML-, 8 ML-, and 3 ML-thick films with the out-of-plane and in-plane magnetic fields taken at different temperatures. The anisotropy in MR could be clearly seen in the thinner films as well, which was pronounced with decreasing temperature while the onset temperature of this anisotropy in MR was determined to be around $T \sim 30$ K for all the samples irrespective of the film thickness.

## D. The basic transport properties of the different-thickness samples

Figure S4a shows the longitudinal resistivity ($\rho_{xx}$) versus temperature curves of the different-thickness samples used in this study, the 30 ML-, 8 ML-, and 3 ML-thick films, respectively. The 30 ML- and 8 ML-thick films showed metallic behavior, whereas the 3 ML-thick film exhibited weakly insulating behavior. Similar evolution from metallic to insulating behavior with reducing thickness was also observed in another itinerant 2D ferromagnet, $Fe_3GeTe_2$,[33] presumably originating from carrier localization effect at the 2D limit. Figure S4b shows the Hall coefficient ($R_H$) versus temperature curves of the same films together with that of bulk $V_5S_8$.[30] The characteristic temperature dependence with the sign change in $R_H$ were commonly observed, indicating that the structures and compositions are essentially the same among those samples.

## E. The detailed AHE data of the different-thickness samples

Figure S5a-c shows the $\rho_{AH}$ versus $\mu_0H$ curves of the different-thickness samples taken at different temperatures. The normal Hall components were subtracted for all the data.

**Additional references**

(28) De Vries, A. B.; Haas, C. *J. Phys. Chem. Solids* **1973**, *34*, 651.

(29) Oka, Y.; Kosuge, K.; Kachi, S. *Phys. Lett.* **1974**, *50A*, 311.

(30) Nozaki, H.; Ishizawa, Y.; Saeki, M.; Nakahira, M. *Phys. Lett.* **1975**, *54A*, 29.

(31) Kitaoka, Y.; Yasuoka, H. *J. Phys. Soc. Jpn.* **1980**, *48*, 1949.

(32) Funahashi, S.; Nozaki, H.; Kawada, I. *J. Phys. Chem. Solids* **1981**, *42*, 1009.

(33) Deng, Y.; Yu, Y.; Song, Y.; Zhang, J.; Wang, N. Z.; Sun, Z.; Yi, Y.; Wu, Y. Z.; Wu, S.; Zhu, J.; Wang, J.; Chen, X. H.; Zhang, Y. *Nature* **2018**, *563*, 94.

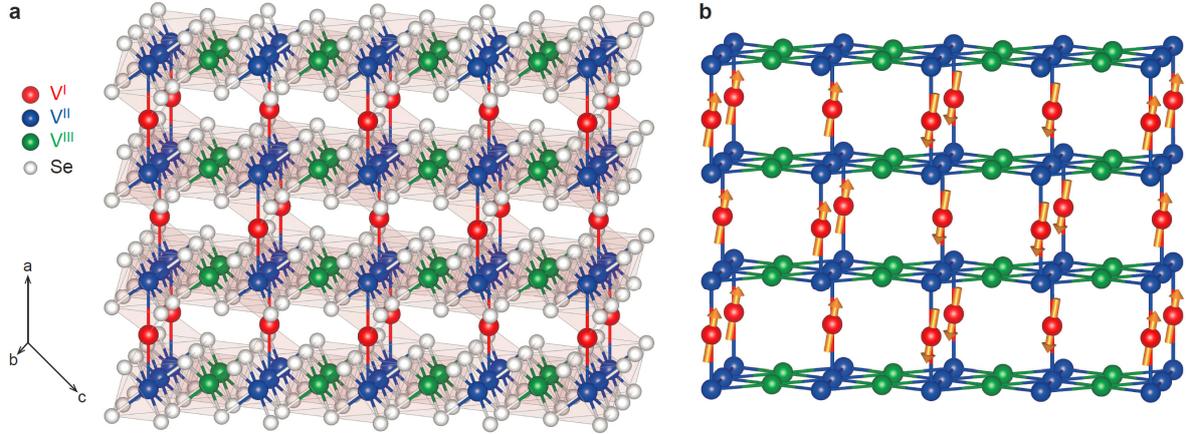

**Figure S1.** (a) The crystal structure of $V_5Se_8$.[18] $V^I$, $V^{II}$ and $V^{III}$ represent the inequivalent V atoms. (b) The magnetic structure of $V_5S_8$ in the AFM phase determined by the neutron diffraction measurements.[32]

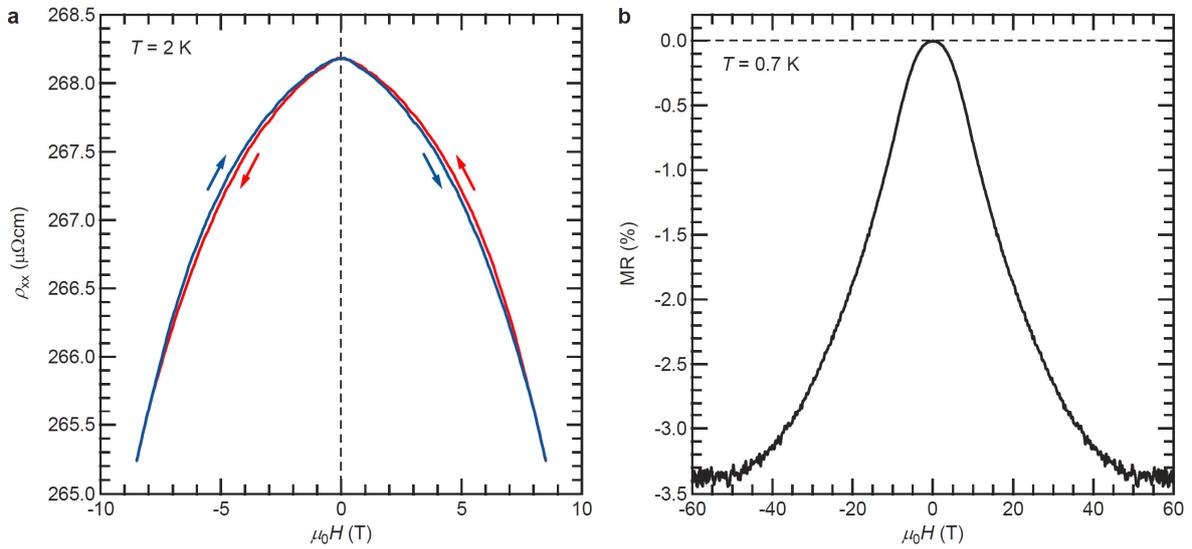

**Figure S2.** (a) The symmetrized MR curve of another 30 ML-thick film measured with the out-of-plane magnetic fields at $T = 2$ K. The magnetic hysteresis loop is most likely originating from the spin-flop transition in the AFM phase. (b) The symmetrized MR curve measured up to $\mu_0H$ ~ 60 T with the out-of-plane magnetic fields at $T = 0.7$ K by the pulsed-field system.

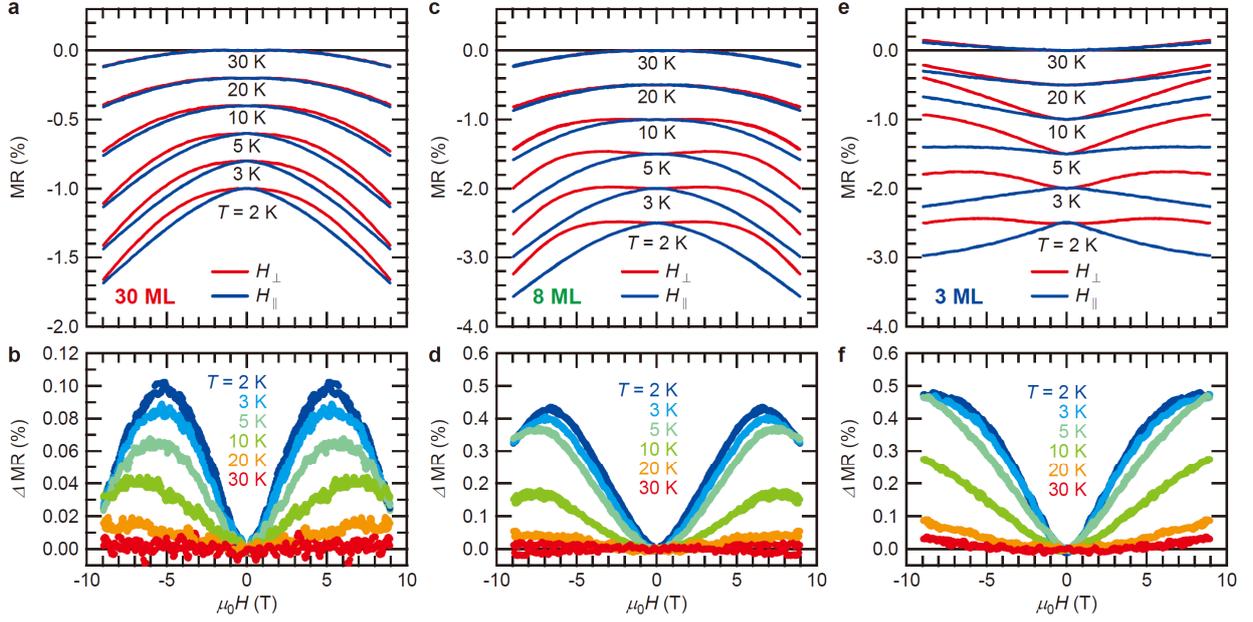

**Figure S3.** The symmetrized MR curves of the (a) 30 ML-, (c) 8 ML-, and (e) 3 ML-thick films, respectively, taken with the out-of-plane and in-plane magnetic fields at different temperatures. The MR value is defined as $[\rho(\mu_0 H)-\rho_0]/\rho_0$. The corresponding difference in MR are shown for the (b) 30 ML-, (d) 8 ML-, and (f) 3 ML-thick films, respectively.

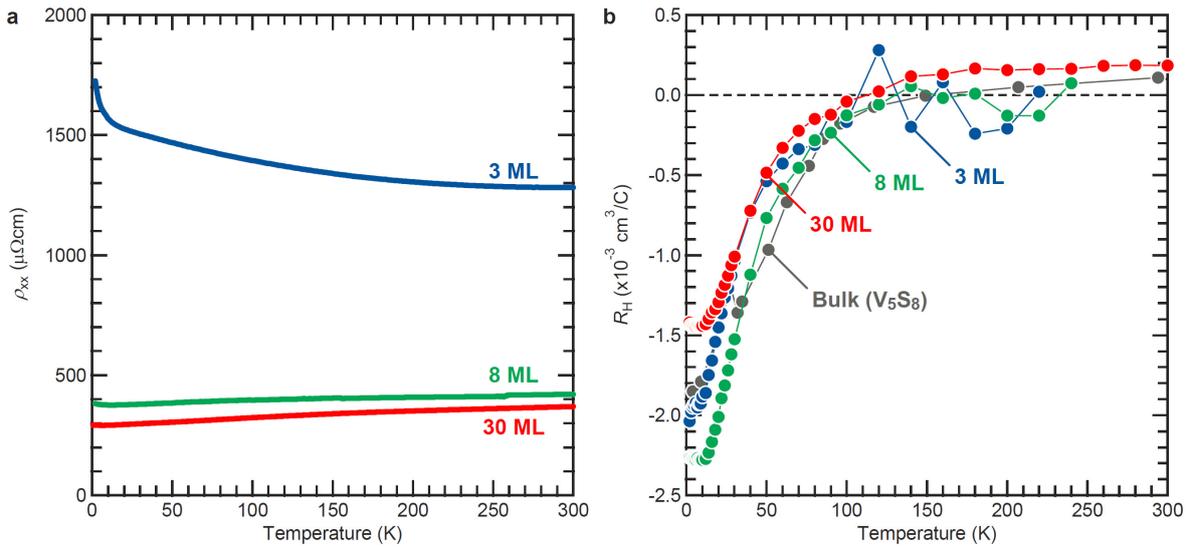

**Figure S4.** (a) The $\rho_{xx}$-$T$ and (b) the $R_H$-$T$ curves of the 30 ML-, 8 ML-, and 3 ML-thick films. The $R_H$-$T$ curve of bulk $V_5S_8$ is also shown.[30]

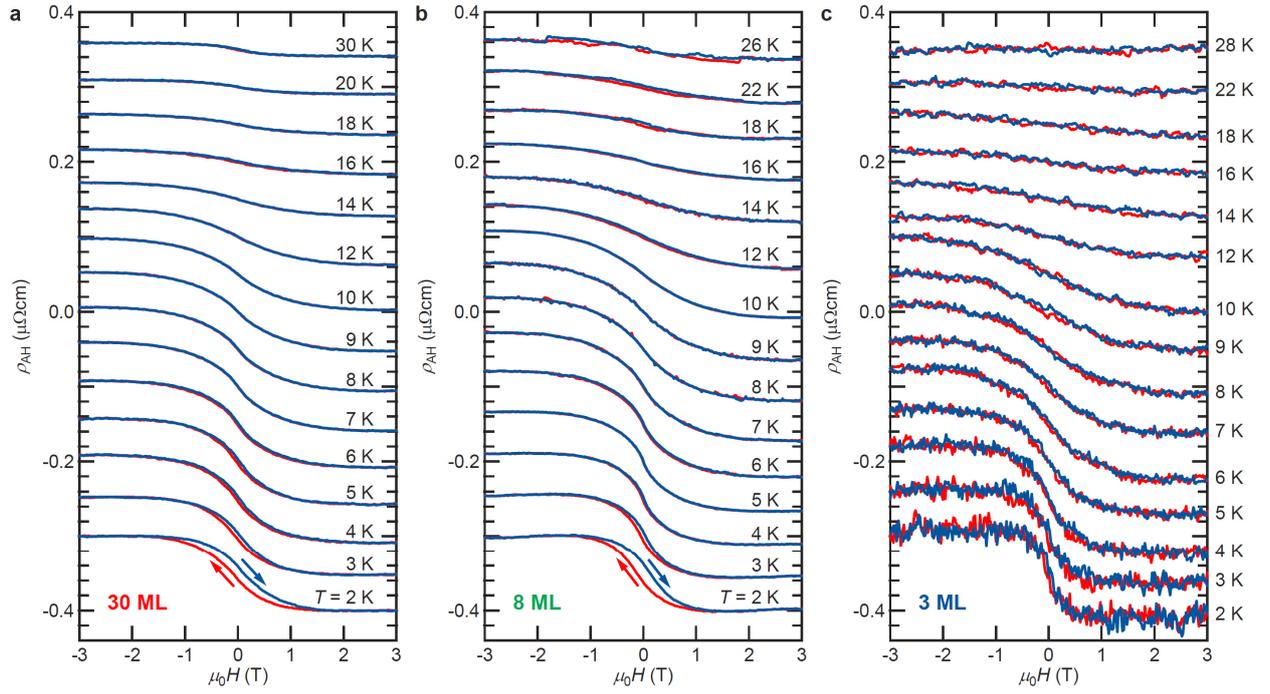

**Figure S5.** The anti-symmetrized $\rho_{AH}$-$\mu_0 H$ curves of the (a) 30 ML-, (b) 8 ML-, and (c) 3 ML-thick films, respectively, taken at different temperatures.